\documentclass{article}
\usepackage{amsfonts}
\usepackage{amsmath}
\usepackage{cite}
\usepackage{appendix}
\usepackage{bm}

\providecommand{\keywords}[1]{\textbf{\text{Keywords:}} #1}

\begin{document}

\title{From the spin-$0$ Duffin-Kemmer-Petiau to the Dirac equation}
\author{Andrzej Okni\'{n}ski \thanks{%
Email: fizao@tu.kielce.pl} \\
%EndAName
Chair of Mathematics and Physics, Politechnika \'{S}wi\c{e}tokrzyska, \\
Al. 1000-lecia PP 7, 25-314 Kielce, Poland}
\maketitle

\begin{abstract}
In the present work a transition from the spin-$0$ Duffin-Kemmer-Petiau
equation to the Dirac equation is described. This transformation occurs when
a crossed field changes into a certain longitudinal field. An experimental
setup to carry out the transition is proposed.
\smallskip 

\noindent{\keywords{relativistic wave equations, boson-to-fermion transition,
supersymmetry}}
\end{abstract}

\section{Introduction}

We have described in our earlier papers connections between the
Duffin-Kemmer-Petiau (DKP) and the Dirac equation, cf. \cite{Okninski2015}
and references therein. In this work a transition from the spin-$0$ DKP
equation to the Dirac equation is described in detail. Although this
transformation is implicit in our previous papers (cf. \cite%
{Okninski2014,Okninski2015a}) we put it in a new perspective and suggest a
possible experiment.

In what follows tensor indices are denoted with Greek letters: $\mu =0,1,2,3$%
. We shall use the following convention for the Minkowski space-time metric
tensor: $g^{\mu \nu }=$ \textrm{diag }$\left( 1,-1,-1,-1\right) $ and we
shall always sum over repeated indices. For example, $a_{\mu }b^{\mu
}=a^{0}b^{0}-\mathbf{a}\cdot \mathbf{b}$. Four-momentum operators are
defined as $p^{\mu }=i\frac{\partial }{\partial x_{\mu }}$ where natural
units have been used: $c=1$, $\hbar =1$. The interaction is introduced via
minimal coupling,

\begin{equation}
p^{\mu }\longrightarrow \pi ^{\mu }=p^{\mu }-qA^{\mu },  \label{minimal}
\end{equation}%
with a four-potential $A^{\mu }$ and a charge $q$. We shall work with
external fields of special configuration, crossed and longitudinal fields,
nonstandard but Lorentz covariant, see \cite{Bagrov2014}. Special matrices
are defined in the appendix \ref{A}.

\section{Duffin-Kemmer-Petiau equations}

Equations considered in this Section, describing spin-$0$ and spin-$1$
bosons, are written as:%
\begin{equation}
\beta _{\mu }\pi ^{\mu }\Psi =m\Psi .  \label{DKP-s0,1}
\end{equation}%
Eq. (\ref{DKP-s0,1}) describes a particle with definite mass if $\beta ^{\mu
}$ matrices obey the commutation relations \cite%
{Tzou1957a,Tzou1957b,Okninski1981,Beckers1995a,Beckers1995b}: 
\begin{equation}
\sum\nolimits_{\lambda ,\mu ,\nu }\beta ^{\lambda }\beta ^{\mu }\beta ^{\nu
}=\sum\nolimits_{\lambda ,\mu ,\nu }g^{\lambda \mu }\beta ^{\nu },
\label{Tzou}
\end{equation}%
where we sum over all permutations of $\lambda ,\mu ,\nu $. It was noticed
in Ref. \cite{Kemmer1939} that $\beta ^{\mu }$ matrices can be realized in
form:%
\begin{equation}
\beta ^{\mu }=\frac{1}{2}\left( \gamma ^{\mu }\otimes I_{4\times
4}+I_{4\times 4}\otimes \gamma ^{\mu }\right) ,  \label{beta}
\end{equation}%
where $g^{\mu \nu }=$ \textrm{diag }$\left( 1,-1,-1,-1\right) $ and $%
I_{4\times 4}$ is a $4\times 4$ unit matrix

It turns out that such $\beta ^{\mu }$ obey simpler but more restrictive
commutation relations  \cite{Kemmer1939,Duffin1938,Petiau1936}:
\begin{equation}
\beta ^{\lambda }\beta ^{\mu }\beta ^{\nu }+\beta ^{\nu }\beta ^{\mu }\beta
^{\lambda }=g^{\lambda \mu }\beta ^{\nu }+g^{\nu \mu }\beta ^{\lambda },
\label{DuffinKemmer}
\end{equation}%
for which Eq.~(\ref{DKP-s0,1}) leads to the Duffin-Kemmer-Petiau (DKP)
theory of spin-$0$ and spin-$1$ mesons. This reducible $16$-dimensional
representation (\ref{beta}) of $\beta ^{\mu }$ matrices (denoted as $\mathbf{%
16}$) can be decomposed as $\mathbf{16}=\mathbf{10}\oplus \mathbf{5}\oplus 
\mathbf{1}$. Explicit formulas for the corresponding $10\times 10$ (spin-$1$
case) and $5\times 5$ (spin-$0$) matrices are given in \cite%
{Beckers1995a,Beckers1995b,Kemmer1939}, while the representation $\mathbf{1}$
is trivial, i.e. all $\beta ^{\mu }=0$.

In the case of representation $\mathbf{5}$ (corresponding to spin-$0$) the
matrices $\beta ^{\mu }$ can be assumed in form (\ref{A1}) to yield: 
\begin{equation}
\left. 
\begin{array}{r}
\pi ^{0}\psi =m\psi ^{0} \\ 
\pi ^{1}\psi =m\psi ^{1} \\ 
\pi ^{2}\psi =m\psi ^{2} \\ 
\pi ^{3}\psi =m\psi ^{3} \\ 
\pi _{0}\psi ^{0}+\pi _{1}\psi ^{1}+\pi _{2}\psi ^{2}+\pi _{3}\psi ^{3}=m\psi%
\end{array}%
\right\} .  \label{DKP1}
\end{equation}

\section{Splitting the $5 \times 5$ DKP equation to a system of two coupled $%
3 \times 3$ equations}

Using relation between a four-vector $\psi ^{\mu }=\left( \psi ^{0},\ 
\bm{\psi }\right) $ and a spinor $\zeta ^{A\dot{B}}$ \cite%
{Berestetskii1974}: 
\begin{equation}
\zeta ^{A\dot{B}}\overset{df}{=}\left( \sigma ^{0}\psi ^{0}+\bm{\sigma }%
\cdot \bm{\psi }\right) ^{A\dot{B}}=\left( 
\begin{array}{cc}
\zeta ^{1\dot{1}} & \zeta ^{1\dot{2}} \\ 
\zeta ^{2\dot{1}} & \zeta ^{2\dot{2}}%
\end{array}%
\right) =\left( 
\begin{array}{cc}
\psi ^{0}+\psi ^{3} & \psi ^{1}-i\psi ^{2} \\ 
\psi ^{1}+i\psi ^{2} & \psi ^{0}-\psi ^{3}%
\end{array}%
\right) ,  \label{4v-spinor}
\end{equation}%
where $\bm{\sigma }=\left( \sigma ^{1},\sigma ^{2},\sigma ^{3}\right) $
are the Pauli matrices (\ref{A2}), $\sigma ^{0}=I_{2\times 2}$, and%
\begin{equation}
\zeta _{1\dot{1}}=\zeta ^{2\dot{2}},~\zeta _{2\dot{2}}=\zeta ^{1\dot{1}%
},~\zeta _{1\dot{2}}=-\zeta ^{2\dot{1}},~\zeta _{2\dot{1}}=-\zeta ^{1\dot{2}%
},  \label{4v-spinor-2}
\end{equation}%
we can rewrite Eq. (\ref{DKP1}) in equivalent form as: 
\begin{equation}
\left. 
\begin{array}{r}
\pi ^{1\dot{1}}\psi =m\psi ^{1\dot{1}} \\ 
\pi ^{2\dot{1}}\psi =m\psi ^{2\dot{1}} \\ 
\pi ^{1\dot{2}}\psi =m\psi ^{1\dot{2}} \\ 
\pi ^{2\dot{2}}\psi =m\psi ^{2\dot{2}} \\ 
\pi _{1\dot{1}}\psi ^{1\dot{1}}+\pi _{2\dot{1}}\psi ^{2\dot{1}}+\pi _{1\dot{2%
}}\psi ^{1\dot{2}}+\pi _{2\dot{2}}\psi ^{2\dot{2}}=2m\psi%
\end{array}%
\right\} ,  \label{DKP2}
\end{equation}%
since $\pi _{1\dot{1}}\psi ^{1\dot{1}}+\pi _{2\dot{1}}\psi ^{2\dot{1}}+\pi
_{1\dot{2}}\psi ^{1\dot{2}}+\pi _{2\dot{2}}\psi ^{2\dot{2}}\equiv 2\left(
\pi _{0}\psi ^{0}+\pi _{1}\psi ^{1}+\pi _{2}\psi ^{2}+\pi _{3}\psi
^{3}\right) $.

Assume that $A^{\mu }$ fulfills: 
\begin{equation}
\left( \pi ^{0}\pi ^{3}-\pi ^{3}\pi ^{0}\right) =\left( \pi ^{2}\pi ^{1}-\pi
^{1}\pi ^{2}\right) =0.  \label{crossed}
\end{equation}%
The condition (\ref{crossed}) is satisfied by the crossed field \cite%
{Bagrov2014}: 
\begin{equation}
\mathbf{E}\cdot \mathbf{n}=\mathbf{H}\cdot \mathbf{n}=\mathbf{E}\cdot 
\mathbf{H}=0,\quad \mathbf{n}=\left[ 0,0,1\right] ,\quad \mid \mathbf{E}\mid
=\mid \mathbf{H}\mid .  \label{A4}
\end{equation}

Now we have: 
\begin{equation}
\pi _{1\dot{1}}\pi ^{1\dot{1}}+\pi _{2\dot{1}}\pi ^{2\dot{1}}=\pi _{1\dot{2}%
}\pi ^{1\dot{2}}+\pi _{2\dot{2}}\pi ^{2\dot{2}}=\pi _{\mu }\pi ^{\mu },
\label{identities1}
\end{equation}%
hence we can also write (splitting the last equation in (\ref{DKP2})):%
\begin{eqnarray}
&&\left. 
\begin{array}{r}
\pi ^{1\dot{1}}\psi =m\psi ^{1\dot{1}} \\ 
\pi ^{2\dot{1}}\psi =m\psi ^{2\dot{1}} \\ 
\pi _{1\dot{1}}\psi ^{1\dot{1}}+\pi _{2\dot{1}}\psi ^{2\dot{1}}=m\psi%
\end{array}%
\right\} ,  \label{DKP1a} \\
&&\left. 
\begin{array}{r}
\pi ^{1\dot{2}}\psi =m\psi ^{1\dot{2}} \\ 
\pi ^{2\dot{2}}\psi =m\psi ^{2\dot{2}} \\ 
\pi _{1\dot{2}}\psi ^{1\dot{2}}+\pi _{2\dot{2}}\psi ^{2\dot{2}}=m\psi%
\end{array}%
\right\} ,  \label{DKP1b}
\end{eqnarray}%
obtaining two $3\times 3$ equations. Eqs. (\ref{DKP1a}), (\ref{DKP1b}),
coupled by component $\psi $, both describe a particle with mass $m$ --
substituting $\psi ^{1\dot{1}}$, $\psi ^{2\dot{1}}$ or $\psi ^{1\dot{2}}$, $%
\psi ^{2\dot{2}}$\ into the corresponding third equations we get $\pi _{\mu
}\pi ^{\mu }\psi =m^{2}\psi $.

\section{From $3\times 3$ equations to the Melosh equations}

Let us substitute $\psi ^{2\dot{1}}$ into the third equation in (\ref{DKP1a}%
). We obtain as in Ref. \cite{Okninski2014}: 
\begin{equation}
\left. 
\begin{array}{r}
\pi ^{1\dot{1}}\psi =m\psi ^{1\dot{1}} \\ 
\pi _{1\dot{1}}\psi ^{1\dot{1}}+\frac{1}{m}\pi _{2\dot{1}}\pi ^{2\dot{1}%
}\psi =m\psi%
\end{array}%
\right\} ,  \label{M1a}
\end{equation}%
or%
\begin{equation}
\left. 
\begin{array}{c}
\left( \pi ^{0}+\pi ^{3}\right) \psi =m\psi ^{1\dot{1}} \\ 
\left( \pi ^{0}-\pi ^{3}\right) \psi ^{1\dot{1}}=m\left( 1+\frac{\pi _{\perp
}^{2}}{m^{2}}\right) \psi%
\end{array}%
\right\} ,  \label{M1b}
\end{equation}%
where $\pi _{\perp }=\left( \pi ^{1},\ \pi ^{2}\right) $ and $\pi _{\perp
}^{2}=\left( \pi ^{1}\right) ^{2}+\left( \pi ^{2}\right) ^{2}$.

Defining a new function%
\begin{equation}
\sqrt{1+\frac{\pi _{\perp }^{2}}{m^{2}}}\thinspace \psi =\varphi ,
\label{def}
\end{equation}%
we get a Melosh-type equation (see Section \ref{DiracMelosh}): 
\begin{equation}
\left. 
\begin{array}{c}
\left( \pi ^{0}+\pi ^{3}\right) \varphi =m\sqrt{1+\frac{\pi _{\perp }^{2}}{%
m^{2}}}\thinspace \psi ^{1\dot{1}} \\ 
\left( \pi ^{0}-\pi ^{3}\right) \psi ^{1\dot{1}}=m\sqrt{1+\frac{\pi _{\perp
}^{2}}{m^{2}}}\thinspace \varphi%
\end{array}%
\right\} .  \label{M1c}
\end{equation}

In the same manner we get from (\ref{DKP1b}): 
\begin{equation}
\left. 
\begin{array}{r}
\pi ^{2\dot{2}}\psi =m\psi ^{2\dot{2}} \\ 
\frac{1}{m}\pi _{1\dot{2}}\pi ^{1\dot{2}}\psi +\pi _{2\dot{2}}\psi ^{2\dot{2}%
}=m\psi%
\end{array}%
\right\} ,  \label{M2a}
\end{equation}%
and, after applying definitions of spinors components and (\ref{def}), we
obtain another Melosh-type equation, discussed in the next Section: 
\begin{equation}
\left. 
\begin{array}{c}
\left( \pi ^{0}-\pi ^{3}\right) \varphi =m\sqrt{1+\frac{\pi _{\perp }^{2}}{%
m^{2}}}\thinspace \psi ^{2\dot{2}} \\ 
\left( \pi ^{0}+\pi ^{3}\right) \psi ^{2\dot{2}}=m\sqrt{1+\frac{\pi _{\perp
}^{2}}{m^{2}}}\thinspace \varphi%
\end{array}%
\right\} .  \label{M2c}
\end{equation}

\section{The Dirac equation and it's Melosh transform}

\label{DiracMelosh}

The Dirac equation, describing spin-$\frac{1}{2}$ elementary particles reads:%
\begin{equation}
\gamma ^{\mu }\pi _{\mu }\Psi =m\Psi ,  \label{Dirac1}
\end{equation}%
where $4\times 4$ matrices $\gamma ^{\mu }$ fulfil relations \cite%
{Dirac1928a,Dirac1928b}:%
\begin{equation}
\gamma ^{\mu }\gamma ^{\nu }+\gamma ^{\mu }\gamma ^{\nu }=2g^{\mu \nu
}I_{4\times 4}.  \label{gamma}
\end{equation}%
In the spinor representation of the Dirac matrices (\ref{A3}) we have $\Psi
=\left( \xi ^{A},\ \eta _{\dot{B}}\right) ^{T}$ \cite{Berestetskii1974},
where $^{T}$ denotes transposition of a matrix. In spinor formalism the
Dirac equation can be written as:%
\begin{equation}
\left. 
\begin{array}{c}
\pi ^{A\dot{B}}\eta _{\dot{B}}=m\xi ^{A}\smallskip \\ 
\pi _{A\dot{B}}\xi ^{A}=m\eta _{\dot{B}}%
\end{array}%
\right\} .  \label{Dirac1S}
\end{equation}%
Multiplying on the left by $\gamma ^{0}$ we get a hamiltonian form of the
Dirac equation:%
\begin{equation}
\pi ^{0}\Psi =\left( \alpha ^{1}\pi ^{1}+\alpha ^{2}\pi ^{2}+\alpha ^{3}\pi
^{3}+\beta m\right) \Psi ,  \label{Dirac2}
\end{equation}%
where $\alpha ^{i}=\gamma ^{0}\gamma ^{i}$, $i=1,2,3$, $\beta =\gamma ^{0}$.
We now assume that the field is longitudinal, i.e. $A^{\mu }=A^{\mu }\left(
x^{0},x^{3}\right) ,\ A^{i}=A^{i}\left( x^{1},x^{2}\right) ,\ \mu =0,3,\
i=1,2$ \cite{Bagrov2014}, so that $\left[ \pi ^{0}\pm \pi ^{3},\pi ^{1}\pm
i\pi ^{2}\right] =0$. Moreover, we demand that $A^{1}=A^{2}=0$ (therefore $%
\pi ^{1,2}=p^{1,2}$ but we keep $\pi^{1,2}$ notation for comparison with the
previous case).

Introducing the fifth anticommuting hermitian matrix $\delta =\alpha
^{1}\alpha ^{2}\alpha ^{3}\beta $ we can transform the Dirac hamiltonian $%
H_{D}=\mathbf{\alpha }\cdot $\textbf{$\pi $}$+\beta m$ by application of a
unitary transformation $U$ \cite{Moss1976}:%
\begin{equation}
U=\frac{1}{\sqrt{2}}\left( \dfrac{\alpha ^{1}\pi ^{1}+\alpha ^{2}\pi
^{2}+\beta m}{\sqrt{\pi _{\perp }^{2}+m^{2}}}+\delta \right) ,  \label{U}
\end{equation}%
arriving at the Melosh form \cite{Melosh1974}:%
\begin{equation}
\pi ^{0}\Phi =\left( -\alpha ^{3}\pi ^{3}+\delta \sqrt{\pi _{\perp
}^{2}+m^{2}}\right) \Phi .  \label{Melosh}
\end{equation}%
More exactly, $\Phi =U\Psi $, $H_{M}=UH_{D}U^{\dagger }$, $H_{M}=-\alpha
^{3}\pi ^{3}+\delta \sqrt{\pi _{\perp }^{2}+m^{2}}$ where $\pi
^{0,3}=p^{0,3}-qA^{0,3}$ and $\pi _{\perp }=\left( \pi ^{1},\pi ^{2}\right)
=\left( p^{1},p^{2}\right) $.

The Melosh hamiltonian involves only two Dirac matrices, $\alpha ^{3}$ and $%
\delta $, so there are several matrices commuting with the Melosh
hamiltonian and hence several possibilities of projecting it onto two
component subspaces, see Section 3 in Ref. \cite{Okninski2014}. For example,
applying projection operators $P_{\pm }=\frac{1}{2}\left( 1\pm i\alpha
^{1}\alpha ^{2}\right) $ we get in the spinor representation of Dirac
matrices:%
\begin{eqnarray}
\pi _{0}\Phi _{A} &=&\left( -\sigma ^{3}\pi ^{3}-\sigma ^{2}\sqrt{\pi
_{\perp }^{2}+m^{2}}\right) \Phi _{A},  \label{Melosh1a} \\
\pi _{0}\Phi _{B} &=&\left( +\sigma ^{3}\pi ^{3}-\sigma ^{2}\sqrt{\pi
_{\perp }^{2}+m^{2}}\right) \Phi _{B},  \label{Melosh1b}
\end{eqnarray}%
where $\Phi _{A}=\left( \Phi _{2},\Phi _{4}\right) ^{T}$ and $\Phi
_{B}=\left( \Phi _{1},\Phi _{3}\right) ^{T}$.

After another unitary transformation, $W=\frac{1}{\sqrt{2}}\left( \sigma
^{1}+\sigma ^{2}\right) \sigma ^{3}$ we obtain: 
\begin{eqnarray}
\pi _{0}W\left( 
\begin{array}{c}
\Phi _{2} \\ 
\Phi _{4}%
\end{array}%
\right) &=&\left( +\sigma ^{3}\pi _{3}+\sigma ^{1}\sqrt{\pi _{1}^{2}+\pi
_{2}^{2}+m^{2}}\right) W\left( 
\begin{array}{c}
\Phi _{2} \\ 
\Phi _{4}%
\end{array}%
\right) ,  \label{Melosh1c} \\
\pi _{0}W\left( 
\begin{array}{c}
\Phi _{1} \\ 
\Phi _{3}%
\end{array}%
\right) &=&\left( -\sigma ^{3}\pi _{3}+\sigma ^{1}\sqrt{\pi _{1}^{2}+\pi
_{2}^{2}+m^{2}}\right) W\left( 
\begin{array}{c}
\Phi _{1} \\ 
\Phi _{3}%
\end{array}%
\right) .  \label{Melosh1d}
\end{eqnarray}

It follows that equations (\ref{M1c}), (\ref{M2c}) and (\ref{Melosh1c}), (%
\ref{Melosh1d}) have the same form but differ in four-potentials.

\section{Outlook and suggestion for experimental setup}

We have described a transition from the DKP equation describing spin-$0$
bosons in a crossed field to the Dirac equation in a longitudinal field,
provided that there is a switchover from crossed to longitudinal field. The
stepping stone of this transformation is the Melosh form: equations (\ref%
{M1c}), (\ref{M2c}) in a crossed field and equations (\ref{Melosh1c}), (\ref%
{Melosh1d}) in longitudinal field. Note, that after the switchover from
crossed to longitudinal field return to the DKP equation is impossible.

This theoretical construction suggests the following experiment. Let a
Bose-Einstein condensate (BEC) of spin-$0$ bosons be prepared in a crossed
field. Then, after a switchover to longitudinal field (cf. \cite%
{Okninski2015a} for a similar idea in a different context), bosons are
expected to transmute to fermions. In different but related case of BEC
obtained in a transition from Bardeen-Cooper-Schrieffer state (BCS) in
ultracold Fermi gases \cite{Zwerger2012} (we assume that Cooper pairs are in
the singlet state) we can expect decay of BEC into BCS (note that Cooper
pairs cannot be pictured as bosons, cf. p. 2 in \cite{Zwerger2012}). %
\appendix{}

\section{Special matrices}

\label{A} \renewcommand{\theequation}{A\arabic{equation}} 
% redefine the command that creates the equation no.
\setcounter{equation}{0} % reset counter 
In the appendix all the necessary definitions of special matrices are
listed. The $5\times 5$ Duffin-Kemmer-Petiau matrices, fulfilling conditions
(\ref{DuffinKemmer}), are: 
\begin{equation}
\begin{array}{ll}
\beta ^{0}=\left( 
\begin{array}{ccccc}
0 & 0 & 0 & 0 & 1 \\ 
0 & 0 & 0 & 0 & 0 \\ 
0 & 0 & 0 & 0 & 0 \\ 
0 & 0 & 0 & 0 & 0 \\ 
1 & 0 & 0 & 0 & 0%
\end{array}%
\right), & \beta ^{1}=\left( 
\begin{array}{ccccc}
0 & 0 & 0 & 0 & 0 \\ 
0 & 0 & 0 & 0 & \hspace{-8pt}-1 \\ 
0 & 0 & 0 & 0 & 0 \\ 
0 & 0 & 0 & 0 & 0 \\ 
0 & 1 & 0 & 0 & 0%
\end{array}%
\right),\medskip \\ 
\beta ^{2}=\left( 
\begin{array}{ccccc}
0 & 0 & 0 & 0 & 0 \\ 
0 & 0 & 0 & 0 & 0 \\ 
0 & 0 & 0 & 0 & \hspace{-8pt}-1 \\ 
0 & 0 & 0 & 0 & 0 \\ 
0 & 0 & 1 & 0 & 0%
\end{array}%
\right), & \beta ^{3}=\left( 
\begin{array}{ccccc}
0 & 0 & 0 & 0 & 0 \\ 
0 & 0 & 0 & 0 & 0 \\ 
0 & 0 & 0 & 0 & 0 \\ 
0 & 0 & 0 & 0 & \hspace{-8pt}-1 \\ 
0 & 0 & 0 & 1 & 0%
\end{array}%
\right),%
\end{array}
\label{A1}
\end{equation}
see also \cite{Kemmer1939} for $\beta^\mu$ matrices defined for different
metric convention.

\noindent The Pauli matrices are assumed in standard form \cite%
{Berestetskii1974}: 
\begin{equation}
\,\sigma ^{1}=\left( 
\begin{array}{cc}
0 & 1 \\ 
1 & 0%
\end{array}%
\right) ,\hspace{8pt}\sigma ^{2}=\left( 
\begin{array}{cc}
0 & \hspace{-8pt}-i \\ 
i & 0%
\end{array}%
\right) ,\hspace{8pt}\sigma ^{3}=\left( 
\begin{array}{cc}
1 & 0 \\ 
0 & \hspace{-8pt}-1%
\end{array}%
\right) .  \label{A2}
\end{equation}%
The Dirac matrices in the spinor representation read \cite{Berestetskii1974}%
: 
\begin{equation}
\begin{array}{ll}
\gamma ^{0}=\left( 
\begin{array}{cccc}
0 & 0 & 1 & 0 \\ 
0 & 0 & 0 & 1 \\ 
1 & 0 & 0 & 0 \\ 
0 & 1 & 0 & 0%
\end{array}%
\right), & \gamma ^{1}=\left( 
\begin{array}{cccc}
0 & 0 & 0 & \hspace{-8pt}-1 \\ 
0 & 0 & \hspace{-8pt}-1 & 0 \\ 
0 & 1 & 0 & 0 \\ 
1 & 0 & 0 & 0%
\end{array}%
\right),\medskip \\ 
\gamma ^{2}=\left( 
\begin{array}{cccc}
0 & 0 & 0 & i \\ 
0 & 0 & \hspace{-8pt}-i & 0 \\ 
0 & \hspace{-8pt}-i & 0 & 0 \\ 
i & 0 & 0 & 0%
\end{array}%
\right), & \gamma ^{3}=\left( 
\begin{array}{cccc}
0 & 0 & \hspace{-8pt}-1 & 0 \\ 
0 & 0 & 0 & 1 \\ 
1 & 0 & 0 & 0 \\ 
0 & \hspace{-8pt}-1 & 0 & 0%
\end{array}%
\right).%
\end{array}
\label{A3}
\end{equation}

\end{document}